\documentclass{svjour2}

\usepackage{graphicx}

\begin{document}

\title{Quasi-normal modes of spherically symmetric black hole spacetimes with cosmic string in a Dirac field}

\author{Sini R         \and
        Nijo Varghese \and V C Kuriakose }

\institute{Sini R
              \email{sini@cusat.ac.in}
               \and
          Nijo Varghese
               \email{nijovarghese@cusat.ac.in}
               \and
         V C Kuriakose
               \email{vck@cusat.ac.in}
                \\
            \emph{Department of Physics, Cochin
University of Science and Technology, Kochi 682022, India.} }

\date{Received: date / Accepted: date}

\maketitle

\begin{abstract}
Dirac equation for a general black hole metric having a cosmic
string is derived. The quasi-normal mode frequencies for
Schwarzschild, RN extremal, SdS and near extremal SdS black hole
space-times with cosmic string perturbed by a massless Dirac field
are obtained using WKB approximation and found that in all these
cases, decay is less in black holes having cosmic string compared to
black holes with out string.
\end{abstract}

\section{Introduction}
\label{intro}Though black holes are natural solutions of Einstein's
general theory of relativity, they are yet to be discovered. Various
theoretical models to account for their discoveries have been
proposed. The study of quasi-normal modes of black holes is one
among them. It is proposed that topological defects, such as
strings, monopoles etc might have been created in the very early
universe\cite{1a}. Among the topological  defects, cosmic string has
been proved to be the most potential one for cosmic structure
formation. The possibility of having strings in the early universe
has been suggested by Kibble in 1976 \cite{1b}. Cosmic strings seem
to be of particular interest because they provide a unique tool to
learn the physics of the very early universe and are considered as a
possible "seed" for galaxy formation \cite{1c,1d} and as a possible
gravitational lens\cite{1e}.

In this paper we consider black holes with the strings\cite{ee}. A
black hole is distinguished by the fact that no information can
escape from within the event horizon and hence the presence of black
holes can be inferred only through indirect methods. The question of
stability of black hole was first treated by Regge and Wheeler
\cite{1f} who investigated linear perturbations of the exterior
Schwarzschild space-time. Further work on this problem \cite{1g} led
to the existence of quasi-normal modes(QNMs) and to the studies of
the response of a black hole to external perturbations. Studies on
perturbations of black holes by gravitational and matter fields have
an important place in black hole physics. Since QNMs depend on black
hole properties such as mass, angular momentum and charge, they
allow a direct way of identifying the space-time parameters.

The QNMs of scalar perturbations around a Schwarzschild black hole
pierced by a cosmic string was done earlier \cite{sch07}. In the
present work we study the influence of cosmic string on the QNMs of
various black hole background space-times which are perturbed by a
massless Dirac field. In section 2, we study the Dirac equation in a
general spherically symmetric space-time with a cosmic sting and its
deduction into a set of second order differential equations. In
Section 3 we evaluate the Dirac quasi-normal frequencies for the
massless case using WKB scheme for Schwarzschild, RN extremal, SdS
and near extremal SdS black hole space-times.

\section{General metric for a black hole with cosmic string space-time perturbed by a Dirac field}
\label{sec:1} The metric describing a spherically symmetric black
hole with a cosmic string can be written as\cite{ee},
\begin{equation}
ds^{2}=-f(r)dt^{2}+\frac{dr^{2}}{f\left( r\right) }+r^{2}d\theta
^{2}+b^{2}r^{2}\sin ^{2}\theta d\phi ^{2}  \label{e}.
\end{equation}
It can be constructed by removing a wedge, which is done by
requiring that the azimuthal angle around the axis runs over the
range $0<\phi'<2\pi b$, with $\phi'=b\phi$ where $\phi$ runs over
zero to $2\pi$. Here $b=1-4\check{\mu }$ with $\check{\mu}$ being
the linear mass density of the string. Following the procedure
adopted in reference \cite{jnwheler}, we develop the Dirac equation
in a general background space-time. We start with the Dirac
equation;
\begin{equation}
\left( \gamma ^{\mu }(\partial _{\mu }-\Gamma _{\mu })+m\right) \Psi
=0 \label{a},
\end{equation}
where $m$ is the the mass of the Dirac field. Here
\begin{equation}
\gamma^{\mu}=g^{\mu\nu}\gamma_{\nu},
\end{equation}
and
\begin{equation}
\gamma_{\nu}=e_{\nu}^{a}\gamma_{a},
\end{equation}
where $\gamma^{a}$ are the Dirac matrices,
\begin{equation}
\gamma_{0}=\left[\begin{array}{cc}
\imath  &0 \\
0 & -\imath%
\end{array}%
\right],\gamma_{i}=\left[\begin{array}{cc}
0 &\sigma_{i} \\
\sigma_{i} &0%
\end{array}%
\right],
\end{equation}
and $\sigma_{i}$ are the Pauli matrices. $ e_{\nu}^{a}$ is the
tetrad given by,
\begin{equation}
e_{t}^{t}=f^{\frac{1}{2}};e_{r}^{r}=\frac{1}{f^{\frac{1}{2}}};e_{\theta
}^{\theta }=r,e_{\phi }^{\phi }=br\sin \theta  \label{111}.
\end{equation}
The inverse of the tetrad $ e_{\nu}^{a }$ is defined by,
\begin{equation}
g^{\mu\nu}=\eta^{ab}e_{a}^{\mu }e_{b}^{\nu },
\end{equation}
with $\eta^{ab}=diag(-1,1,1,1)$, the Minkowski metric. The spin
connection $\Gamma_{\mu} $ is given by
\begin{equation}
\Gamma _{\mu}=-\frac{1}{2}[\gamma_{a},\gamma_{b}] e_{\nu
}^{a}e^{b\nu}_{ ;\mu} \label{b},
\end{equation}
where $ e^{b\nu}_{ ;\mu} =\partial _{\mu }e^{b\nu }+\Gamma _{\kappa
\mu }^{\nu }e^{b\kappa } $ is the covariant derivative of $e^{b\nu
}$. The spin connections for the above metric are obtained as,
\begin{eqnarray}
\Gamma _{t} &=&\frac{1}{4}\gamma _{1}\gamma _{0}\frac{\partial
f\left( r\right) }{\partial r},
\label{113} \\
\Gamma _{r} &=&0, \\
\Gamma _{\theta } &=&\frac{1}{2}\gamma _{1}\gamma
_{2}f^{\frac{1}{2}}\left(
r\right) , \\
\Gamma _{\phi } &=&\frac{1}{2}\gamma _{2}\gamma _{3}b\cos \theta +\frac{1}{2}%
\gamma_{1}\gamma _{3}b\sin \theta f^{\frac{1}{2}}\left( r\right).
\end{eqnarray}
Substituting the spin connections in Eq. (\ref{a}) we will get ,
\begin{equation}
[\frac{-\gamma _{0}}{f^{\frac{1}{2}}}\frac{\partial }{\partial t}
+\gamma _{1}f^{\frac{1}{2}}( \frac{\partial }{\partial r}+\frac{1}{r}+%
\frac{1}{4{f( r) }}{\frac{\partial f}{\partial r} }) +
\frac{\gamma_{2}}{r}(\frac{\partial }{\partial \theta }+\frac{1}{2}\cot\theta)+\frac{\gamma_{3}}{%
br\sin \theta }\frac{\partial }{\partial \phi }+m]
\Psi(t,r,\theta,\phi) =0 \label{114a},
\end{equation}
Using the transformation $\Psi(t,r,\theta,\phi)=\frac{\exp(-\imath
Et)}{rf^{\frac{1}{4}}{\sin\theta}^{\frac{1}{2}}}\chi(r,\theta,\phi)$,
Eq.(\ref{114a}) becomes,
\begin{eqnarray}
[\frac{-E}{f^{\frac{1}{2}}}\frac{\partial }{\partial t}%
+\frac{1}{\imath}\gamma_{0}\gamma _{1}f^{\frac{1}{2}}(
\frac{\partial }{\partial r})  +
\frac{\gamma_{1}}{\imath r}\gamma_{1}\gamma_{0}(\gamma _{2}\frac{\partial }{\partial \theta }&+&\nonumber\\ \frac{\gamma_{3}}{%
b\sin \theta }\frac{\partial }{\partial \phi })-\imath \gamma_{0}m]
\chi(r,\theta,\phi) =0 \label{114}.
\end{eqnarray}
Dirac equation can be separated out into radial and angular parts by
the following substitution,
\begin{equation}
\chi(r,\theta,\phi)=R(r)\Omega(\theta,\phi)\label{215}.
\end{equation}
The angular momentum operator is introduced as, \cite{jnwheler}
\begin{equation}
\mathbf{K}_{\left( b\right) }=-\imath\gamma _{1}\gamma_{0}\left(
\gamma _{2}\partial _{\theta }+\gamma _{3}\left( b\sin \theta
\right) ^{-1}\partial _{\phi }\right)   \label{116},
\end{equation}
such that,
\begin{equation}
\mathbf{K}_{ (b)}\Omega(\theta,\phi)=k_{b}\Omega(\theta,\phi),
\label{a117}
\end{equation}
where $k_{b}=\frac{k}{b}$ are the eigenvalues of $\mathbf{K}_{(b)}$.
Here k is a positive or a negative nonzero integer with
$l=|k+\frac{1}{2}|-\frac{1}{2}$, where $l$ is the total orbital
angular momentum. The cosmic string presence is codified in the
eigenvalues of the angular momentum operator\cite{1j}. Substituting
Eqs.(\ref{215}) and (\ref{a117}) in Eq.(\ref{114}), we will get
radial equation which contains $\gamma _{0}$ and $\gamma _{1}$. As
$\gamma _{0}$ and $\gamma _{1}$ can be represented by $2\times2$
matrices, we write the radial factor $R(r)$ by a two component
spinor notation,
\begin{equation}
R(r)=\left[\begin{array}{c}
F \\
G%
\end{array}\right].%
\end{equation}
Then the radial equation in F an G are given by,

\begin{equation}
f\frac{dG}{dr}+f^{\frac{1}{2}}\frac{k_{b}}{r}G+f^{\frac{1}{2}}mF=E F
\label{i},
\end{equation}
\begin{equation}
f\frac{dF}{dr}-f^{\frac{1}{2}}\frac{k_{b}}{r}F+f^{\frac{1}{2}}mG=-E
G \label{j}.
\end{equation}
Introducing a co-ordinate change as,
\begin{equation}
dr_{\ast }=\frac{dr}{f}  \label{117},
\end{equation}%
Eq.(\ref{i}) and Eq.(\ref{j}) can be combined into a single equation
as,
\begin{equation}
\frac{\partial }{\partial r_{\ast }}\left[
\begin{array}{c}
G \\
F%
\end{array}%
\right] +f^{\frac{1}{2}}\left[
\begin{array}{cc}
\frac{k_{b}}{r} & m \\
m & -\frac{k_{b}}{r}%
\end{array}%
\right] \left[
\begin{array}{c}
G \\
F%
\end{array}%
\right] =\left[
\begin{array}{cc}
0 & E \\
-E  & 0%
\end{array}%
\right] \left[
\begin{array}{c}
G \\
F%
\end{array}%
\right]   \label{118}.
\end{equation}
Defining,
\begin{equation}
\left[
\begin{array}{c}
\hat{G} \\
\hat{F}
\end{array}%
\right] =\left[
\begin{array}{cc}
\cos \frac{\theta}{2} & -\sin \frac{\theta}{2} \\
\sin \frac{\theta}{2} & \cos \frac{\theta}{2}%
\end{array}%
\right] \left[
\begin{array}{c}
G \\
F
\end{array}%
\right]   \label{119},
\end{equation}
where for positive value of k,
\begin{equation}
\theta =\tan ^{-1}(\frac{mr}{|k_{b}|})  \label{120}.
\end{equation}%
Eq.(\ref{118})  now becomes,
\begin{equation}
\frac{\partial }{\partial r_{\ast }}\left[
\begin{array}{c}
\hat{G} \\
\hat{F}%
\end{array}%
\right]
+f^{\frac{1}{2}}\sqrt{\left(\frac{k_{b}}{r}\right)^{2}+m^{2}}\left[
\begin{array}{cc}
1 & 0 \\
0 & -1%
\end{array}%
\right] \left[
\begin{array}{c}
\hat{G} \\
\hat{F}%
\end{array}%
\right] =-E \left[ 1+\frac{1}{2E }\frac{fm|k_{b}|}{{%
{k_{b}}^{2}+m^{2}r^{2}}}\right] \left[
\begin{array}{cc}
0 & -1 \\
1 & 0%
\end{array}%
\right] \left[
\begin{array}{c}
\hat{G} \\
\hat{F}%
\end{array}%
\right]   \label{121}.
\end{equation}
By making another change of the variable;
\begin{equation}
d\hat{r}_{\ast }=\frac{dr_{\ast }}{\left[ 1+\frac{1}{2E }\frac{fm|k_{b}|}{{%
k_{b}^{2}+m^{2}r^{2}}}\right] }  \label{122}.
\end{equation}
Eq.(\ref{121}) can be simplified to,
\begin{equation}
\frac{\partial }{\partial \hat{r}_{\ast }}\left[
\begin{array}{c}
\hat{G} \\
\hat{F}%
\end{array}%
\right] +\frac{f^{\frac{1}{2}}\sqrt{\left(\frac{k_{b}}{r}\right)^{2}+m^{2}}}{\left[ 1+\frac{1}{%
2E }\frac{fm|k_{b}|}{{k_{b}^{2}+m^{2}r^{2}}}\right] }\left[
\begin{array}{cc}
1 & 0 \\
0 & -1%
\end{array}%
\right] \left[
\begin{array}{c}
\hat{G}\\
\hat{F}%
\end{array}%
\right] =E \left[
\begin{array}{cc}
0 & 1 \\
-1 & 0%
\end{array}%
\right] \left[
\begin{array}{c}
\hat{G} \\
\hat{F}%
\end{array}%
\right]   \label{123},
\end{equation}
i.e,
\begin{equation}
\frac{\partial }{\partial \hat{r}_{\ast }}\left[
\begin{array}{c}
\hat{G} \\
\hat{F}%
\end{array}%
\right] +W\left[
\begin{array}{c}
\hat{G} \\
-\hat{F}%
\end{array}%
\right] =E \left[
\begin{array}{c}
\hat{F} \\
-\hat{G}%
\end{array}%
\right]   \label{124},
\end{equation}
where
\begin{equation}
W=\frac{f^{\frac{1}{2}}\sqrt{\left(\frac{k_{b}}{r}\right)^{2}+m^{2}}}{\left[ 1+\frac{1}{2E }%
\frac{fm|k_{b}|}{{k_{b}^{2}+m^{2}r^{2}}}\right] }  \label{125}.
\end{equation}%
Thus from Eq.(\ref{124}), we will get two coupled equations for
$\hat{G}$ and $\hat{F}$ which are given bellow,
\begin{equation}
-\frac{\partial ^{2}\hat{F}}{\partial \hat{r}_{\ast
}^{2}}+V_{1}\hat{F}=E ^{2}\hat{F} \label{126},
\end{equation}
\begin{equation}
-\frac{\partial ^{2}\hat{G}}{\partial \hat{r}_{\ast
}^{2}}+V_{2}\hat{G}=E^{2}\hat{G} \label{127},
\end{equation}%
where
\begin{equation}
V_{1,2}=\pm \frac{\partial W}{\partial \hat{r}_{\ast }}+W^{2}
\label{128}.
\end{equation}
From Eq. (\ref{126}) and Eq. (\ref{127}), we can evaluate the
quasi-normal mode frequencies for various black hole space-times.
Here $V_{1}$ and $V_{2}$ are the super symmetric partners derived
from the same super potential $W$ \cite{aa} and these potentials
give same spectra of quasi-normal mode frequencies.
\section{Quasi-normal mode frequencies}
\label{sec:2}
 We shall now evaluate the quasi-normal frequencies for various black hole space-times perturbed
by a massless Dirac field using WKB approximation. As $V_{1}$ and
$V_{2}$ give same spectra of quasi-normal mode frequencies we avoid
the subscripts and write $V$ for the potential function. Thus, for
massless case the equation for the potential given by Eq.(\ref{128})
becomes,

\begin{equation}
V= f\frac{\partial \left( f^{\frac{1}{2}}\frac{k_{b}}{r}\right) }{\partial r}%
+f\left(\frac{k_{b}}{r}\right)^{2}  \label{134}.
\end{equation}

\subsection{Schwarzschild black hole}
\begin{figure}[h]
\center
\includegraphics[width=5.5cm]{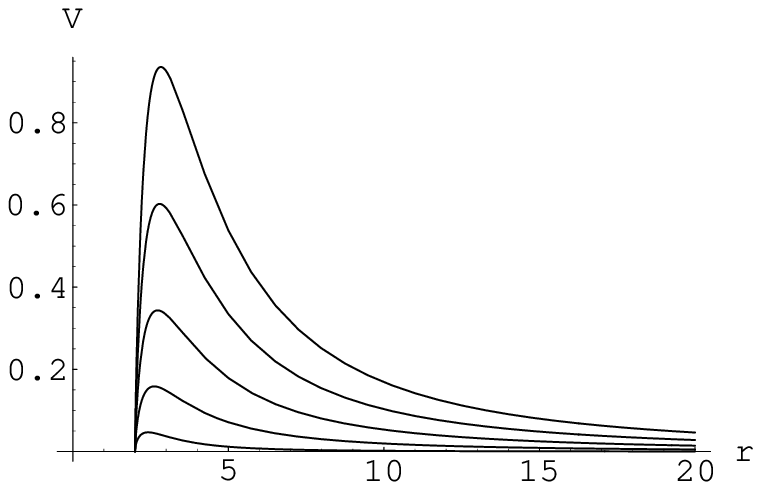}
\includegraphics[width=5.5cm]{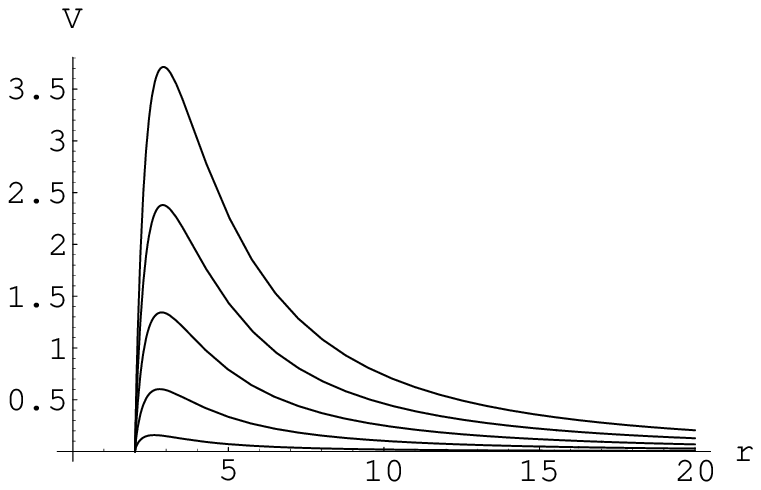}
\includegraphics[width=5.5cm]{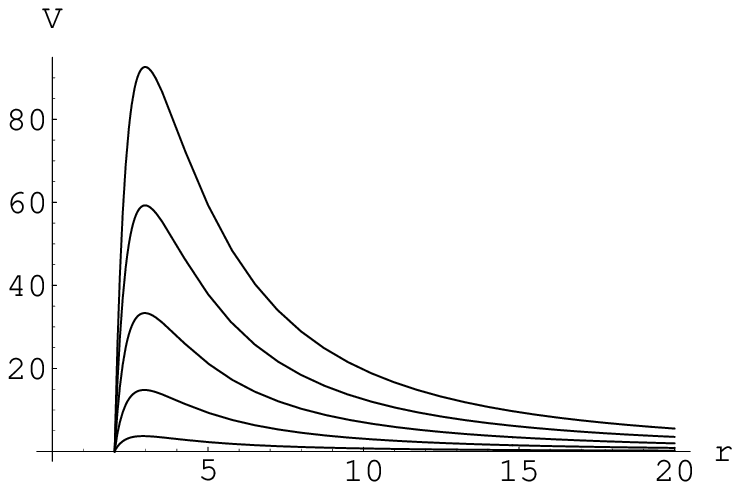}
\caption{Potential versus r for Schwarzschild black hole with cosmic
string in Dirac field is plotted for different values of $b$ ($b=1,
0.5, 0.1$) for $k=0$ to $k=5$. }\label{graph1}
\end{figure}

We first consider the most simple black hole, viz., the
Schwarzschild black hole for which,
\begin{equation}
\label{171}f(r)=\left( 1-\frac{2M}{r}\right).
\end{equation}
Substituting the above $f$ in Eq.(\ref{134}) we get,
\begin{equation}
V=\left( 1-\frac{2M}{r}\right) \frac{\partial }{\partial r}%
\left( \left( 1-\frac{2M}{r}\right)^{\frac{1}{2}} \frac{k_{b}}{r}\right) +\left( 1-\frac{2M}{r}%
\right) \left(\frac{k_{b}}{r}\right)^{2}  \label{172},
\end{equation}
where $M$ is the mass of the Schwarzschild black hole. The effective
potential $V$ which depends on the absolute value of $k_{b}$, is in
the form of a barrier. The peak of the barrier gets higher and
higher as $|k|$ increases for fixed $b$ values. We repeat the
calculation for different $b$ values ($b=1, 0.5, 0.1$), and find
that the height of the potential increases with $b$ values
decreasing. i.e, the presence of the cosmic string, causes an
increase in the height of the potential (Fig. \ref{graph1}).

To evaluate the quasi-normal mode frequencies, we use the WKB
approximation \cite{bb,cc,dd}.  This method can be accurate for both
the real and imaginary parts of the frequencies for low lying modes
with $n\prec k$, where $n$ is the mode number and $k$ is the angular
momentum quantum number. The formula for complex quasi-normal mode
frequencies E in the WKB approximation, carried out to third order
is,
\begin{equation}
E^{2}=\left[ V_{0}+\left( -2V_{0}^{^{\prime \prime }}\right) ^{\frac{1}{2}%
}\Lambda \right] -i\left( n+\frac{1}{2}\right) \left(
-2V_{0}^{^{\prime \prime }}\right) ^{\frac{1}{2}}\left( 1+\Omega
\right),   \label{139}
\end{equation}
where
\begin{equation}
\Lambda =\frac 1{(-2V_0^{^{\prime \prime }})^{1/2}}\left[ \frac
18\left( \frac{V_0^{(4)}}{V_0^{^{\prime \prime }}}\right) \left(
\frac 14+\alpha
^2\right) -\frac 1{288}\left( \frac{V_0^{^{\prime \prime \prime }}}{%
V_0^{^{\prime \prime }}}\right) ^2\left( 7+60\alpha ^2\right)
\right],\label{140}
\end{equation}

\begin{equation}
\Omega =\frac 1{(-2V_0^{^{\prime \prime }})}\left\{
\begin{array}{c}
\frac 5{6912}\left( \frac{V_0^{^{\prime \prime \prime
}}}{V_0^{^{\prime \prime }}}\right) ^4\left( 77+188\alpha ^2\right)
-\frac 1{384}\left( \frac{V_0^{^{\prime \prime \prime
}2}V_0^{^{(4)}}}{V_0^{^{\prime \prime }3}}\right) \left(
51+100\alpha ^2\right) \\ +\frac 1{2304}\left(
\frac{V_0^{(4)}}{V_0^{^{\prime \prime }}}\right) ^2\left(
67+68\alpha ^2\right) +\frac 1{288}\left( \frac{V_0^{^{\prime \prime
\prime }}V_0^{^{(5)}}}{V_0^{^{\prime \prime }2}}\right) \left(
19+28\alpha
^2\right) \\ -\frac 1{288}\left( \frac{V_0^{^{(6)}}}{V_0^{^{\prime \prime }}}%
\right) \left( 5+4\alpha ^2\right)
\end{array}
\right\}.\label{141}
\end{equation}
Here
\begin{equation}
\alpha =n+\frac 12, n=\{
\begin{array}{c}
{0 , 1 , 2 , ..., Re(E)}>0 \\ {-1 , -2 , -3, ..., RE(E)}< 0,
\end{array}\label{142}
\end{equation}
\begin{equation}
V_0^{^{(n)}}=\frac{d^nV}{dr_{*}^n}\mid _{r_{*}=r_{*}\left( r_{\max
}\right) },\label{143}
\end{equation}

where the values of $n$ lie in the range: $0\leq n<k$. Plugging the
effective potential in Eq.(\ref{172}) in to the formula given above,
we obtain the complex quasi-normal mode frequencies for
Schwarzschild black hole having cosmic string perturbed by a
massless Dirac field. The values of $Re(E)$ and $Im(E)$ calculated
for different values of $b$ are given in Table \ref {tab:1}.

\begin{table}
\caption{Quasi-normal frequencies of Schwarzschild black hole with
cosmic string.}
 \label{tab:1}
 \center
\begin{tabular}{|c|c|c|c|c|}
\hline $b$ & $k$ & $n$ & $ReE$ & $ImE$\\ \hline 0.1 & 1 &
0 & 1.92351 & -0.0962324 \\ \hline & 2 & 0 & 3.84851 & -0.0962269 \\
\hline &  & 1 & 3.84584 & -0.288768 \\ \hline & 3 & 0 & 5.77297 &
-0.0966667 \\ \hline &  & 1 & 5.77131 & -0.290031 \\ \hline &  & 2 &
5.768 & -0.483485 \\ \hline & 4 & 0 & 7.69776 & -0.0962255  \\
\hline &  & 1 & 7.69642 & -0.288698 \\ \hline &  & 2 & 7.69375 &
-0.481236 \\ \hline &  & 3 & 7.68976 & -0.673882 \\ \hline & 5 & 0 &
9.62231 & -0.0962253 \\ \hline &  & 1 & 9.62124 & -0.28869  \\
\hline &  & 2 & 9.6191 & -0.481196  \\ \hline &  & 3 & 9.6159 &
-0.673772 \\ \hline &  & 4 & 9.61165 & -0.866445 \\ \hline
 0.5 & 1 &
0 & 0.378627 & -0.0965424 \\ \hline & 2 & 0 & 0.767194 & -0.096276
\\ \hline &  & 1 & 0.753957 & 0.291048 \\ \hline & 3 & 0 & 1.15303 &
-0.0962463 \\ \hline &  & 1 & 1.14416 & -0.289714 \\ \hline &  & 2 &
1.12741 & -0.485759 \\ \hline & 4 & 0 & 1.53836 & -0.0962367  \\
\hline &  & 1 & 1.5317 & -0.289257 \\ \hline &  & 2 & 1.51881 &
-0.483803 \\ \hline &  & 3 & 1.50043 & -0.680554  \\ \hline & 5 & 0
& 1.92351 & -0.0962324  \\ \hline &  & 1 & 1.91818 & -0.289047 \\
\hline &  & 2 & 1.90773 & - 0.482862  \\ \hline &  & 3 & 1.89259 &
- 0.678198 \\ \hline &  & 4 & 1.87327 & - 0.87539  \\ \hline
 1 & 1 & 0
& 0.176452 & - 0.100109 \\ \hline & 2 & 0 & 0.378627 & - 0.0965424
\\ \hline &  & 1 & 0.353604 & - 0.298746  \\ \hline & 3 & 0 &
0.573685 &  - 0.0963242 \\ \hline &  & 1 & 0.556185 & - 0.292981  \\
\hline &  & 2 & 0.527289 & - 0.497187 \\ \hline & 4 & 0 & 0.767194 &
- 0.096276 \\ \hline &  & 1 & 0.753957 & - 0.291048  \\ \hline &  &
2 & 0.73045 &  - 0.49088  \\ \hline &  & 3 & 0.699918 &  - 0.695711
\\ \hline & 5 & 0 & 0.960215 &  - 0.0962564  \\ \hline &  & 1 &
0.949593 &  - 0.290179  \\ \hline &  & 2 & 0.929979 & - 0.487634 \\
\hline &  & 3 &
0.903578 & - 0.689241  \\ \hline &  & 4 & 0.872052 & - 0.894412  \\
\hline
\end{tabular}
\end{table}

\begin{figure}[h]
\center
\includegraphics[width=8cm]{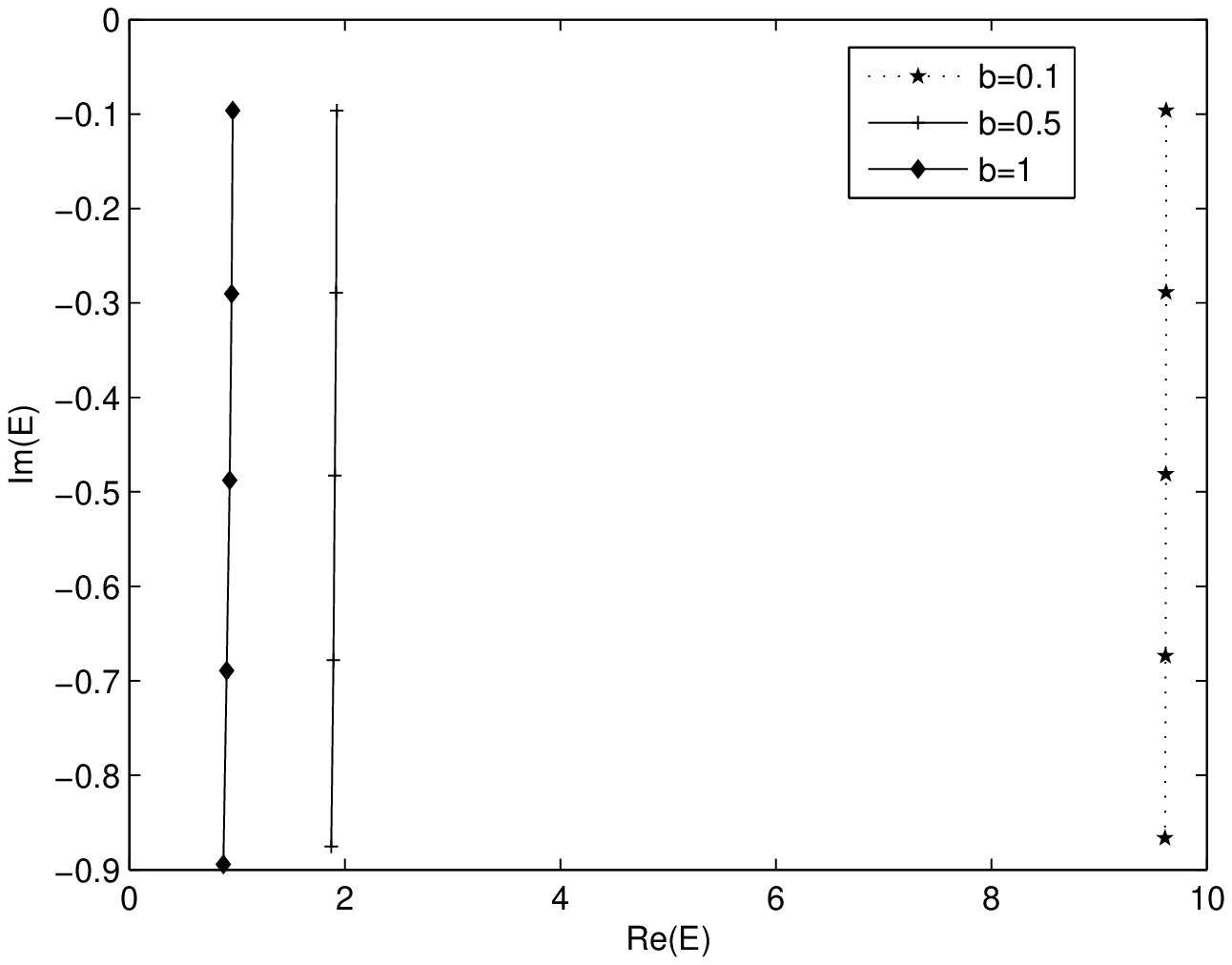}
\caption{Quasi-normal modes of Schwarzschild black hole with cosmic
string perturbed by a  massless Dirac field is plotted for different
values of $b$ ($b=0.1,0.5$, and $1$) for $k=5$.}\label{graph2}
\end{figure}

For a fixed $b$ value, $Re(E)$ decreases as the mode number $n$
increases for the same angular momentum quantum number $k$ and
$|Im(E)|$ increases with $n$. This indicates that quasi-normal modes
with higher mode numbers decay faster than the low-lying one. The
variation of mode frequencies for a fixed $k$ and changing the $b$
values are shown in Fig. \ref{graph2}. When the cosmic string effect
is large, i.e when $b$ is small, $Re(E)$ increases and $|Im(E)|$
decreases for a fixed $k$ [Table \ref {tab:1}]. This implies that
the decay is less in the case of Schwarzschild black hole having
cosmic string compared to the case of black hole without string. For
$b=1$ case, we obtain the same results given in Ref.\cite{ff} where
the quasi-normal modes of Schwarzschild black hole perturbed by a
massless Dirac field was calculated.

\subsection{RN extremal black hole}
\begin{figure}[h]
\center
\includegraphics[width=5.5cm]{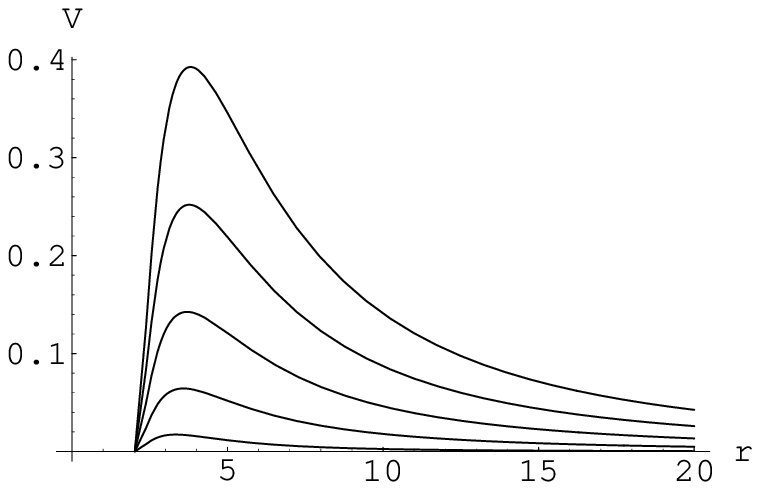}
\includegraphics[width=5.5cm]{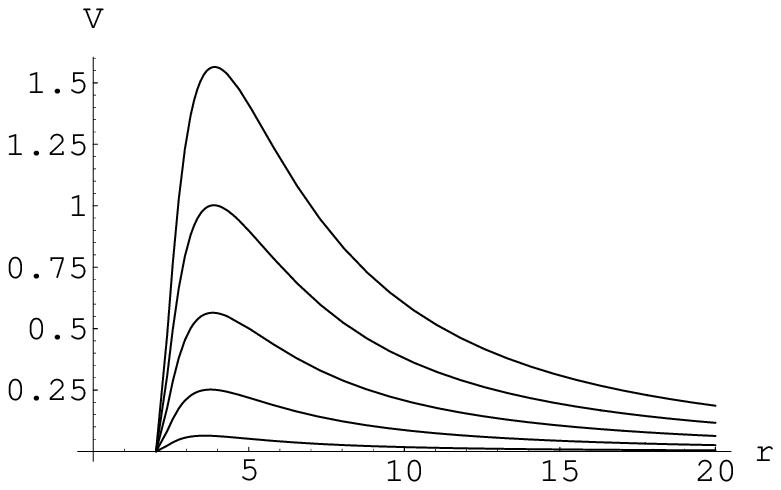}
\includegraphics[width=5.5cm]{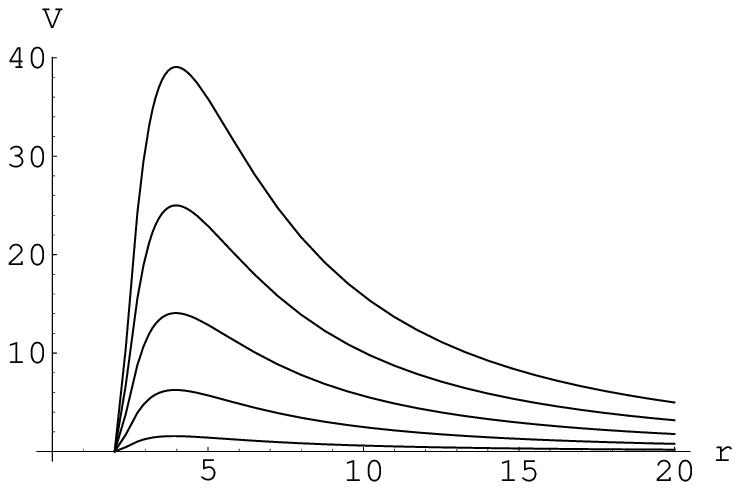}
\caption{Potential versus r for RN extremal black hole with cosmic
string in Dirac field is plotted for different values of $b$ ($b=1,
0.5, 0.1$) for $k=0$ to $k=5$. }\label{graph3}
\end{figure}

Now we will consider RN extremal black hole for which,
\begin{equation}
\label{131}f(r)=\left( 1-\frac{r_{0}}{r}\right) ^{2}.
\end{equation}
Substituting the above $f$ in Eq.(\ref{134}) we will get,
\begin{equation}
V=\left( 1-\frac{r_{0}}{r}\right) ^{2}\frac{\partial }{\partial r}%
\left( \left( 1-\frac{r_{0}}{r}\right) \frac{k_{b}}{r}\right) +\left( 1-\frac{r_{0}}{r}%
\right) ^{2}\left(\frac{k_{b}}{r}\right)^{2}  \label{136}.
\end{equation}
Here the barrier potential $V$ depends on the absolute value of
$k_{b}$ and for a fixed $b$ value, the peak of the barrier gets
higher and higher as $|k|$ increases. We now take 3 different values
for $b(b=1, 0.5, 0.1)$ and from Fig. \ref{graph3} it is found that
the presence of the cosmic string causes an increase in the peak of
the potential.

When the effective potential  given in Eq.(\ref{136}) is substituted
in Eq.(\ref{139}), we will get the complex quasi-normal mode
frequencies of RN extremal black hole perturbed by a massless Dirac
field [Table \ref{tab:2}].

\begin{figure}[h]
\center
\includegraphics[width=8cm]{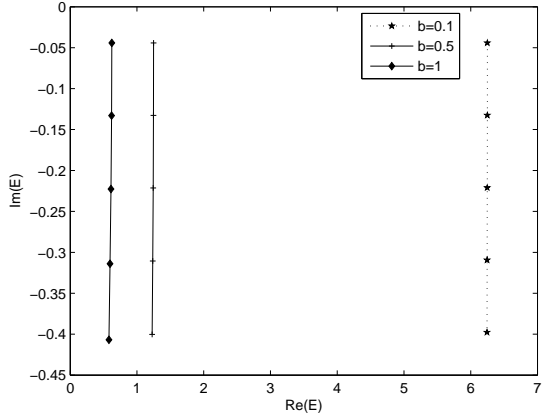}
\caption{Quasi-normal modes of RN extremal black hole with cosmic
string perturbed by a massless Dirac field is plotted for different
values of $b$ ($b=0.1,0.5$, and $1$) for $k=5$.}\label{graph4}
\end{figure}

\begin{table}
\caption{Quasi-normal modes of RN extremal black hole having Cosmic
string} \label{tab:2} \center
       % Give a unique label
\begin{tabular}{|c|c|c|c|c|}
\hline $b$ & $k$ & $n$ & $ReE$ & $ImE$ \\ \hline 0.1 & 1 &
0 & 1.24941 & -0.0441915 \\ \hline & 2 & 0 & 2.49971 & -0.0441935 \\
\hline &  & 1 & 2.49853 & -0.132603 \\ \hline & 3 & 0 & 3.7498 &
-0.0441939 \\ \hline &  & 1 & 3.74902 & -0.132591 \\ \hline &  & 2 &
3.74746 & -0.221018 \\ \hline & 4 & 0 & 4.99985 & -0.044194 \\
\hline &  & 1 & 4.99927 & -0.132588 \\ \hline &  & 2 & 4.9981 &
-0.220998 \\ \hline &  & 3 & 4.99634 & -0.309435 \\ \hline & 5 & 0 &
6.24988 & -0.0441941 \\ \hline &  & 1 & 6.24941 & -0.132586 \\
\hline &  & 2 & 6.24848 & -0.220988 \\ \hline &  & 3 & 6.24707 &
-0.309408 \\ \hline &  & 4 & 6.2452 & -0.397852 \\ \hline
 0.5 & 1 & 0 & 0.246679 &
-0.044139 \\ \hline & 2 & 0 & 0.498474 & -0.0441761
\\ \hline &  & 1 & 0.492516 & -0.133115 \\ \hline & 3 & 0 & 0.749004
& -0.0441865 \\ \hline &  & 1 & 0.745065 & -0.132812 \\ \hline &  &
2 & 0.737355 & -0.222156 \\ \hline & 4 & 0 & 0.999259 & -0.0441899
\\ \hline &  & 1 & 0.996315 & -0.13271 \\ \hline &  & 2 & 0.990498 &
-0.221638 \\ \hline &  & 3 & 0.981942 & -0.311209 \\ \hline & 5 & 0
& 1.24941 & -0.0441915 \\ \hline &  & 1 & 1.24706 & -0.132664 \\
\hline &  & 2 & 1.24239 & -0.221398 \\ \hline &  & 3 & 1.23548 &
-0.310554 \\ \hline &  & 4 & 1.22643 & -0.400269 \\ \hline
 1 & 1 & 0
& 0.117481 & -0.0444736 \\ \hline & 2 & 0 & 0.246679 & -0.0441391 \\
\hline &  & 1 & 0.234639 & -0.135043 \\ \hline & 3 & 0 & 0.37291 &
-0.0441622 \\ \hline &  & 1 & 0.364905 & -0.133573 \\ \hline &  & 2
& 0.350195 & -0.225651 \\ \hline & 4 & 0 & 0.498474 & -0.0441761 \\
\hline &  & 1 & 0.492516 & -0.133115 \\ \hline &  & 2 & 0.481162 &
-0.481162 \\ \hline &  & 3 & 0.465287 & -0.316232 \\ \hline & 5 & 0
& 0.623796 & -0.0441828 \\ \hline &  & 1 & 0.619053 & -0.132916 \\
\hline &  & 2 & 0.609859 & -0.222676 \\ \hline &  & 3 & 0.596707 &
-0.313914 \\ \hline &  & 4 & 0.580161 & -0.406822  \\ \hline
\end{tabular}
\end{table}
Here also we find, for a fixed $b$ value, $Re(E)$ decreases while
$|Im(E)|$ increases with the mode number $n$ increasing for the same
angular momentum eigenvalue $k$. This means that quasi-normal modes
with higher mode numbers decay faster than the low-lying ones. For a
fixed $k$, the variation of mode frequencies with $b$ values are
shown in Fig. \ref{graph4}. When $b<1$, i.e, when the cosmic string
is present,  $Re(E)$ increases and  $|Im(E)|$ decreases for a fixed
$k$, compared to the $b=1$ case [Table \ref {tab:2}]. Thus compared
to RN extremal black hole, the decay is less in the case of RN
extremal black hole having cosmic string.

\subsection{Schwarzschild-de Sitter black hole}
\begin{figure}[h]
\center
\includegraphics[width=5.5cm]{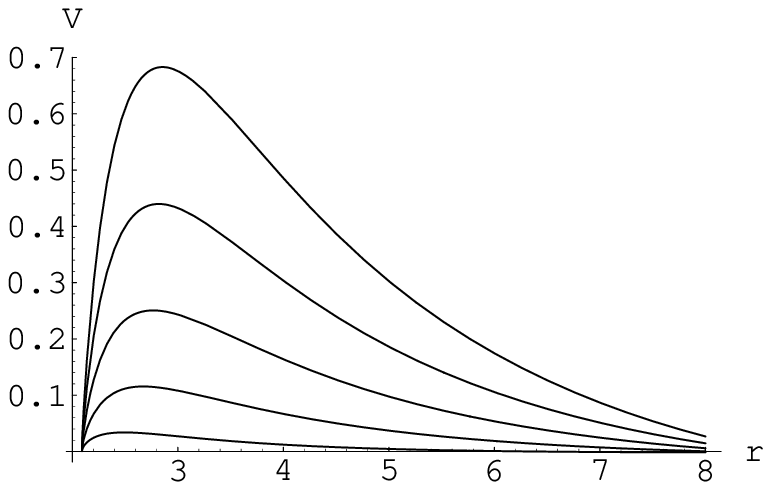}
\includegraphics[width=5.5cm]{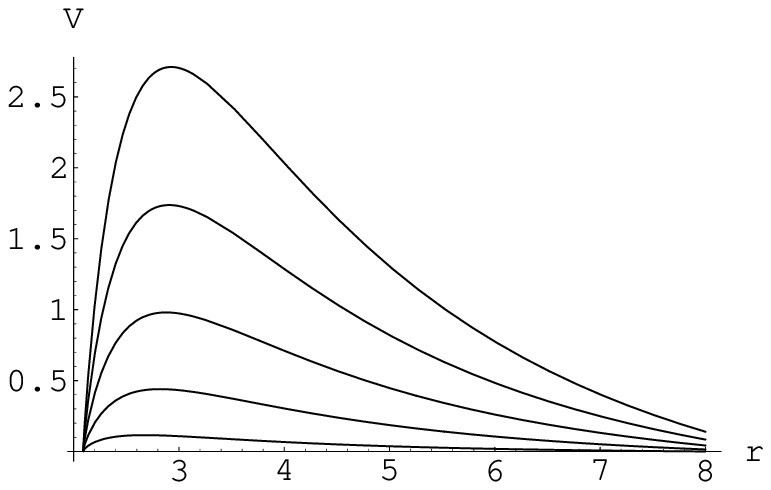}
\includegraphics[width=5.5cm]{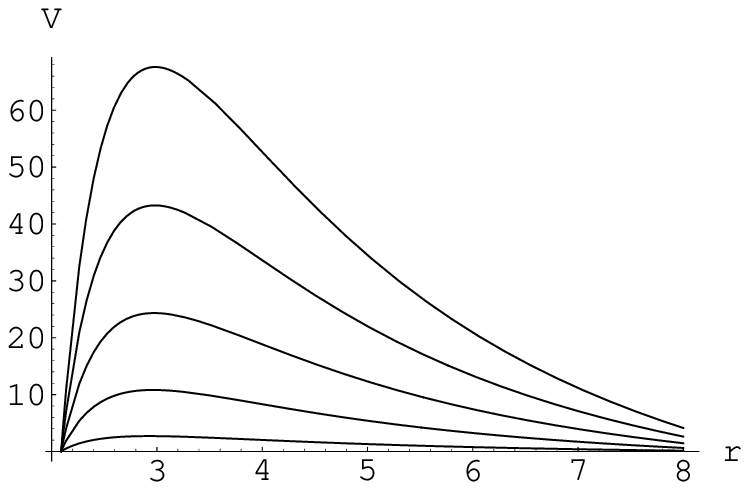}
\caption{Potential versus r for SdS black hole with cosmic string in
Dirac field is plotted for different values of $b$ ($b=1, 0.5, 0.1$)
for $k=0$ to $k=5$. }\label{graph5}
\end{figure}

We will now take the case of SdS black hole and see how the
quasi-normal modes are effected when the cosmic string pierces a SdS
black hole. The metric is,
\begin{equation}
\label{129}f(r)=1-\frac{2M}r-\frac{r_{}^2}{a^2},
\end{equation}
where $M$ denotes the black hole mass and $a^2=\frac 3\Lambda $,
$\Lambda $ being the cosmological constant. The space-time possesses
two horizons: the black-hole horizon at $r=r_{b}$ and the
cosmological horizon at $r=r_{c}$. The function $f$ has zeros at
$r_{b}$, $r_{c}$  and $r_{0}=-(r_{b}+r_{c})$. Substituting the above
$f$ in Eq.(\ref{134}), we will get
\begin{equation}
V=\left( 1-\frac{2M}{r}-\frac{r^{2}}{a^{2}}\right) \frac{\partial
}{\partial r}\left( \left( 1-\frac{2M}{r}-\frac{r^{2}}{a^{2}}\right) ^{\frac{%
1}{2}}\frac{k_{b}}{r}\right) +\left(
1-\frac{2M}{r}-\frac{r^{2}}{a^{2}}\right)\left(\frac{k_{b}}{r}\right)^{2}
\label{137}.
\end{equation}
We consider three values of $b$ ($b=1, 0.5, 0.1$) and from Fig.
\ref{graph5} we can see that the height of the potential increases
with b decreasing.

By substituting Eq.(\ref{137}) in Eq.(\ref{139}), we obtain the
complex quasi-normal modes for SdS black hole with cosmic string
perturbed by a massless Dirac field. Quasi-normal modes of SdS black
hole with cosmic string for various $b$ values are shown in Table
\ref{tab:3}.
\begin{table}
\caption{Quasi-normal frequencies of SdS black hole with cosmic
string}\small \label{tab:3} \center
\begin{tabular}{|c|c|c|c|c|}
\hline $b$ & $k$ & $n$ & $ReE$ & $ImE$ \\ \hline 0.1 & 1 &
0 & 1.62881 & -0.193788 \\ \hline & 2 & 0 & 3.28115 & -0.193023 \\
\hline &  & 1 & 3.26217 & -0.57955 \\ \hline & 3 & 0 & 4.92798 &
-0.192659 \\ \hline &  & 1 & 4.91537 & -0.578183  \\ \hline &  & 2 &
4.89017 & -0.964325  \\ \hline & 4 & 0 & 6.57352 & -0.192473 \\
\hline &  & 1 & 6.56408 & -0.577532 \\ \hline &  & 2 & 6.5452 &
-0.962932 \\ \hline &  & 3 & 6.51691 & -1.3489 \\ \hline & 5 & 0 &
8.21856 & -0.192351 \\ \hline &  & 1 & 8.211 & -0.577126 \\ \hline &
& 2 & 8.19591 & -0.962118 \\ \hline &  & 3 & 8.17328 & -1.34747 \\
\hline &  & 4 & 8.14313 & -1.73333 \\ \hline  0.5 & 1 &
0 & 0.184451 & -0.179973 \\ \hline & 2 & 0 & 0.611882 & -0.191711 \\
\hline &  & 1 & 0.506702 & -0.602245 \\ \hline & 3 & 0 & 0.959022 &
-0.193731 \\ \hline &  & 1 & 0.891965 & -0.589723 \\ \hline &  & 2 &
0.768017 & -1.00536 \\ \hline & 4 & 0 & 1.29563 & -0.19389 \\ \hline
&  & 1 & 1.2466 & -0.585655 \\ \hline &  & 2 & 1.15204 & -0.98806 \\
\hline &  & 3 & 1.01702 & -1.40482 \\ \hline & 5 & 0 & 1.62881 &
-0.193748 \\ \hline &  & 1 & 1.59011 & -0.583538  \\ \hline &  & 2 &
1.51421 & -0.979819 \\ \hline &  & 3 & 1.40371 & -1.38574 \\ \hline
&  & 4 & 1.26152 & -1.80321 \\ \hline  1 & 1 & 0
& 0.118178 & 0.329652 \\ \hline & 2 & 0 & 0.18445 & -0.179973 \\
\hline &  & 1 & 0.0172463 & -0.686957 \\ \hline & 3 & 0 & 0.423791 &
-0.187411 \\ \hline &  & 1 & 0.283273 & -0.623849 \\ \hline &  & 2 &
0.0808308 & -1.10559 \\ \hline & 4 & 0 & 0.611882 & -0.191711 \\
\hline &  & 1 & 0.5067 & -0.602246 \\ \hline &  & 2 & 0.335459 &
-1.05152 \\ \hline &  & 3 & 0.108969 & -1.52834 \\ \hline & 5 & 0 &
0.787788 & -0.193209 \\ \hline &  & 1 & 0.705661 & -0.593841 \\
\hline &  & 2 & 0.560701 & -1.02226 \\ \hline &  & 3 & 0.365525 &
-1.47586 \\ \hline &  & 4 & 0.124334 & -1.95241  \\ \hline
\end{tabular}
\end{table}

\begin{figure}[h]
\center
\includegraphics[width=8cm]{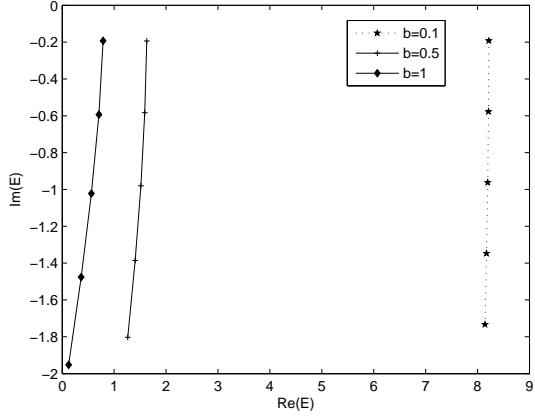}
\caption{Quasi-normal modes of SdS black hole with cosmic string
perturbed by a massless Dirac field is plotted for different values
of $b$ ($b=0.1,0.5$, and $1$) for $k=5$. }\label{graph6}
\end{figure}

Here for a fixed $b$ value, $Re(E)$ and $|Im(E)|$ show similar
behavior as those of Schwarzschild and RN extremal black holes
having cosmic string. The variation of mode frequencies for a fixed
$k$ with different $b$ values are shown in Fig. \ref{graph6}. From
Table \ref {tab:3}, we can see that the behavior of $|Im(E)|$ for
$n=0$ mode is different from the behavior of $|Im(E)|$ for non zero
values of $n$.  From this table we can see that the decay is less in
the case of SdS black hole having cosmic string.

\subsection{Near extremal Schwarzschild-de Sitter black hole.}
\begin{figure}[h]
\center
\includegraphics[width=5.5cm]{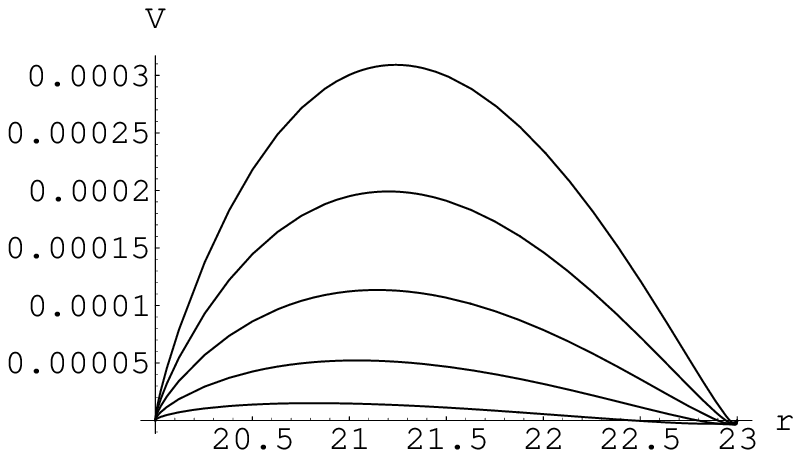}
\includegraphics[width=5.5cm]{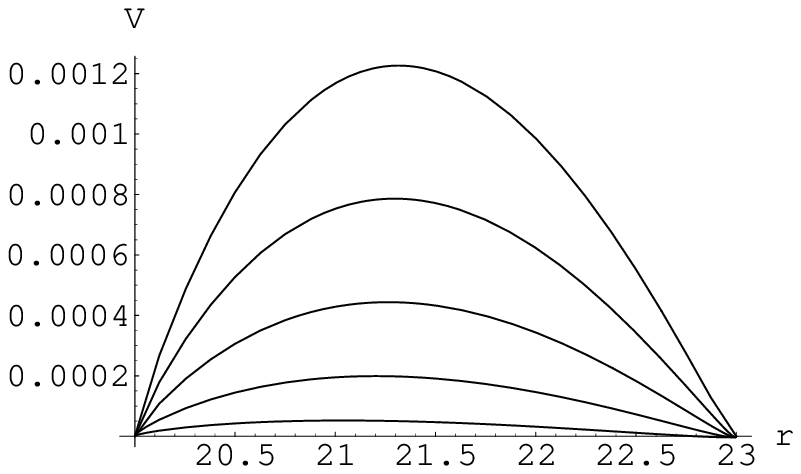}
\includegraphics[width=5.5cm]{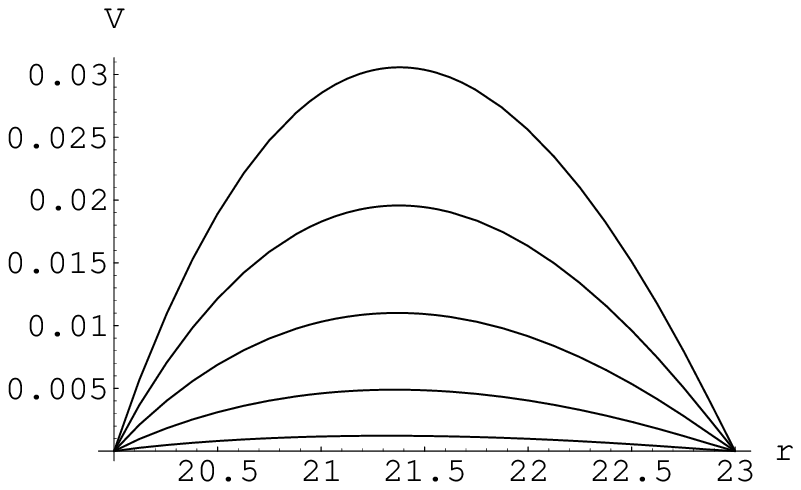}
\caption{Potential versus r for SdS near extremal black hole with
cosmic string in Dirac field is plotted for different values of $b$
($b=1, 0.5, 0.1$) for $k=0$ to $k=5$. }\label{graph7}
\end{figure}

As a last example, we consider near extremal SdS black hole, which
is defined as the space-time for which the cosmological horizon
$r_{c}$ is very close to black hole horizon $r_{b}$, i.e,
$\frac{r_{c}-r_{b}}{r_{b}}<<1$. For this space-time, one can make
the following approximations,
\begin{equation}
r_{0}\sim 2r_{b};a^{2}\sim 3r_{b}^{2};M\sim \frac{r_{b}}{3}
\label{1122}.
\end{equation}
Furthermore, since $r$ is constrained to vary between $r_{b}$ and
$r_{c}$, we get $r-r_{0}\sim r_{b}-r_{0}\sim 3r_{b}$ and thus
\begin{equation}
f\sim \frac{\left( r-r_{b}\right) \left( r_{c}-r\right)
}{r_{b}^{2}}. \label{1123}
\end{equation}
Substituting  Eq.(\ref{1123}) in Eq.(\ref{134}), we get,
\begin{equation}
V= \frac{\left( r-r_{b}\right) \left( r_{c}-r\right) }{r_{b}^{2}}%
\frac{\partial }{\partial r}\left( \left( \frac{\left(
r-r_{b}\right) \left( r_{c}-r\right) }{r_{b}^{2}}\right)
^{\frac{1}{2}}\frac{k_{b}}{r}\right)
+\left( \frac{\left( r-r_{b}\right) \left( r_{c}-r\right) }{r_{b}^{2}}%
\right) \left( \frac{k_{b}}{r}\right) ^{2} . \label{1124}
\end{equation}
From Fig. \ref{graph7} we can see that the height of the potential
increases when the effect of cosmic string increases.

By substituting Eq.(\ref{1124}) in Eq.(\ref{139}) we obtain the
quasi-normal modes of near extremal SdS black hole with cosmic
string perturbed by a massless Dirac field. For a near extremal SdS
black hole also the quasi-normal frequencies are obtained for
different values of $b$ [Table \ref{tab:4}].
\begin{table}
\caption{Quasi-normal frequencies of SdS  near extremal black hole
with cosmic string} \label{tab:4} \center
\begin{tabular}{|c|c|c|c|c|}
\hline $b$ & $k$ & $n$ & $ReE$ & $ImE$ \\ \hline 0.1 & 1 & 0 &
0.0349689 &  - 0.001875 \\ \hline & 2 & 0 & 0.0699379 & - 0.001875  \\
\hline &  & 1 & 0.0699379 &  - 0.005625  \\ \hline & 3 & 0 &
0.104907 &  - 0.001875 \\ \hline &  & 1 & 0.104907 &  - 0.005625  \\
\hline &  & 2 & 0.104907 & - 0.009375 \\ \hline & 4 & 0 & 0.139876 &
- 0.001875 \\ \hline &  & 1 & 0.139876 & - 0.005625  \\ \hline &  &
2 & 0.139876 &  - 0.009375 \\ \hline &  & 3 & 0.139876 &  - 0.013125
\\ \hline & 5 & 0 & 0.174845 & -0.001875 \\ \hline &  & 1 & 0.174845
&  - 0.005625 \\ \hline &  & 2 & 0.174845 &  - 0.009375  \\ \hline &
& 3 & 0.174845 &  - 0.013125  \\ \hline &  & 4 & 0.174845 &  -
0.016875  \\ \hline 0.5 & 1 & 0 & 0.0069931 &  0.00187594 \\ \hline
& 2 & 0 & 0.0139875 & - 0.00187507 \\ \hline &  & 1 & 0.0139876 &  -
0.00562519 \\ \hline & 3 & 0 & 0.0209814 &   -0.00187501 \\ \hline &
& 1 & 0.0209814 &  -0.00562504  \\ \hline &  & 2 & 0.0209814
& -0.00937506 \\ \hline & 4 & 0 & 0.0279751 & - 0.001875  \\
\hline & & 1 & 0.0279751 & - 0.00562501  \\ \hline &  & 2 &
0.0279752 & - 0.00937502  \\ \hline &  & 3 & 0.0279752 &  - 0.013125
\\ \hline & 5 & 0 & 0.0349689 & - 0.001875 \\ \hline &  & 1 &
0.0349689 & - 0.00562501 \\ \hline &  & 2 & 0.0349689 &  -
0.00937501  \\ \hline & & 3 & 0.0349689 & - 0.013125  \\ \hline &  &
4 & 0.0349689 & - 0.016875  \\ \hline 1 & 1 & 0 & 0.00348641 &  -
0.00188365 \\ \hline & 2 & 0 & 0.0069931 & - 0.00187594 \\ \hline &
& 1 & 0.00699432 &
- 0.00562683  \\ \hline & 3 & 0 & 0.0104906 & - 0.00187521  \\
\hline &  & 1 & 0.0104908 &  - 0.00562551 \\ \hline &  & 2 &
0.0104911 & - 0.00937561  \\ \hline & 4 & 0 & 0.0139875] & -
0.00187507 \\ \hline &  & 1 & 0.0139876 &  - 0.00562519  \\ \hline &
& 2 & 0.0139877 &  - 0.00937525 \\ \hline &  & 3 & 0.0139878 &  -
0.0131253 \\ \hline & 5 & 0 & 0.0174845 & - 0.00187503 \\ \hline & &
1 & 0.0174845 & - 0.00562508  \\ \hline &  & 2 & 0.0174845 & -
0.00937512 \\ \hline &  & 3 & 0.0174846 &  - 0.0131251 \\ \hline & &
4 & 0.0174846 &  - 0.0168751 \\ \hline
\end{tabular}
\end{table}

\begin{figure}[h]
\center
\includegraphics[width=8cm]{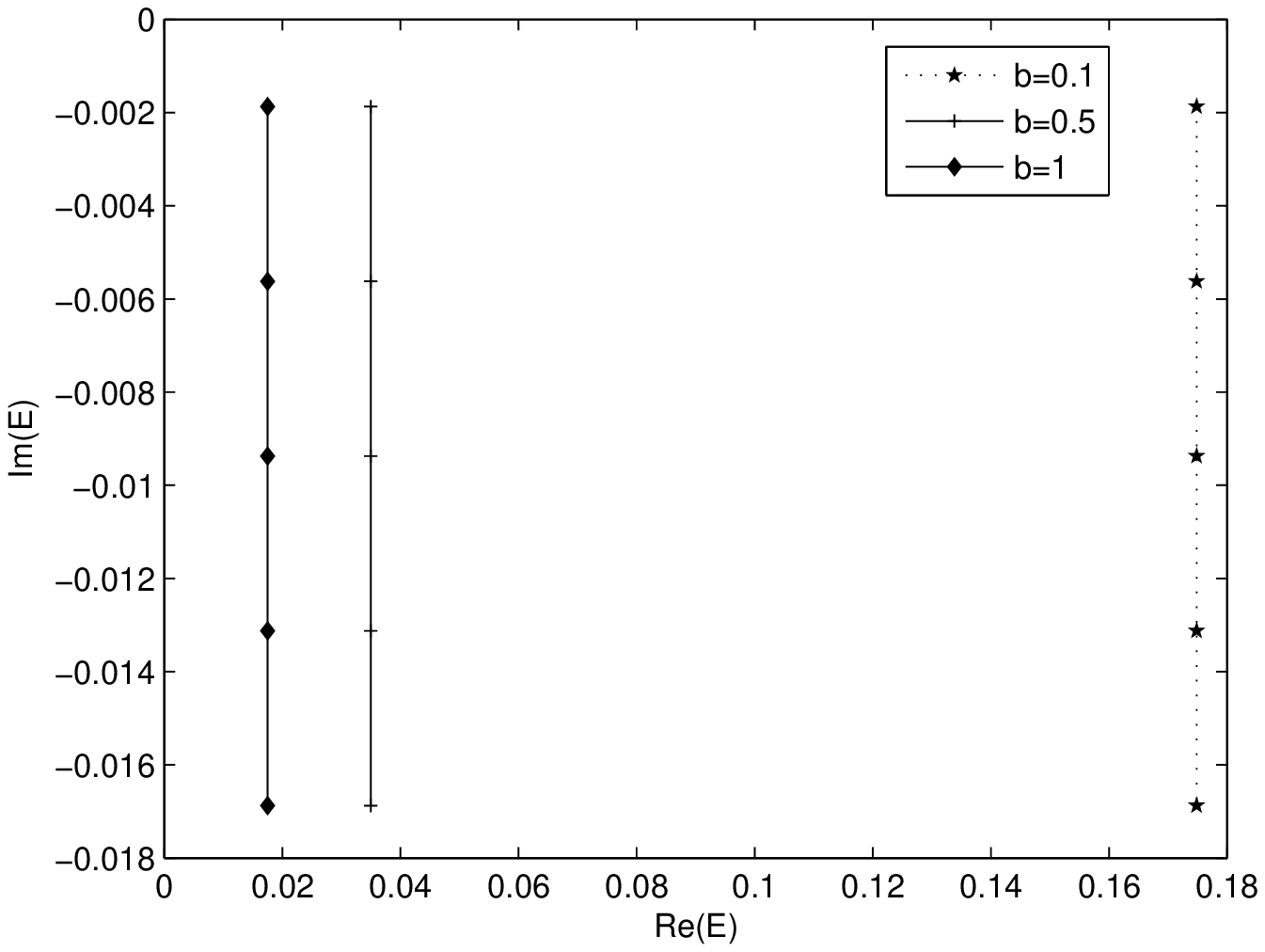}
\caption{Quasi-normal modes of near extremal SdS  black hole with
cosmic string perturbed by a massless Dirac field is plotted for
different values of $b$ ($b=0.1,0.5$, and $1$) for $k=5$.
}\label{graph8}
\end{figure}

We can see that for a fixed $b$ value, $Re(E)$ remains same and
$|Im(E)|$ increases as the mode number n increases for the same $k$
value. The behavior of mode frequencies by changing the $b$ values
for a fixed $k$ are shown in Fig. \ref{graph8}. When $b$ is small,
i.e, when the effect of cosmic string is high, $Re(E)$ increases
while $|Im(E)|$ have almost same but with very small decreases for a
mode of fixed $k$ value [Table \ref {tab:4}]. The decay is less in
the case of near extremal SdS black hole having cosmic string.

\section{Conclusion}
 We have evaluated the quasi-normal mode frequencies for Schwarzschild, RN extremal, SdS and near
extremal SdS black hole space-times having cosmic string perturbed
by a massless Dirac field.  In all these cases, we have found that
quasi-normal modes with higher mode numbers decay faster than the
low-lying ones. We have also found that, when the effect of cosmic
string is high, $|Im(E)|$ decreases while $Re(E)$ increases for
fixed $k$ implying that the decay is less when cosmic string is
present.

\begin{acknowledgements}
SR and VCK are thankful to U.G.C, Government of India for financial
support in the form of a project. VCK wishes to acknowledge
Associateship of IUCAA, Pune. India.
\end{acknowledgements}

\end{document}